\documentclass[sigconf,natbib=false,authorversion]{acmart}

\usepackage[utf8]{inputenc}
\usepackage{subcaption}
\usepackage{balance}

\usepackage{siunitx}
\DeclareSIUnit\sample{S}
\DeclareSIUnit[group-minimum-digits=5]\usd{USD}

\copyrightyear{2023}
\acmYear{2023}
\setcopyright{licensedothergov}\acmConference[CCS '23]{Proceedings of the 2023 ACM SIGSAC Conference on Computer and Communications Security}{November 26--30, 2023}{Copenhagen, Denmark}
\acmBooktitle{Proceedings of the 2023 ACM SIGSAC Conference on Computer and Communications Security (CCS '23), November 26--30, 2023, Copenhagen, Denmark}
\acmPrice{15.00}
\acmDOI{10.1145/3576915.3623135}
\acmISBN{979-8-4007-0050-7/23/11}


\usepackage[style=acmnumeric,citestyle=numeric-comp,backend=biber,sorting=none]{biblatex}
\title{Watch This Space: Securing Satellite Communication through Resilient Transmitter Fingerprinting}

\newcommand{\sysname}[0]{\textsc{SatIQ}}

\author{Joshua Smailes}
\affiliation{%
    \institution{University of Oxford}
    \city{Oxford}
    \country{UK}
}
\email{joshua.smailes@cs.ox.ac.uk}

\author{Sebastian K{\"o}hler}
\affiliation{%
    \institution{University of Oxford}
    \city{Oxford}
    \country{UK}
}
\email{sebastian.kohler@cs.ox.ac.uk}

\author{Simon Birnbach}
\affiliation{%
    \institution{University of Oxford}
    \city{Oxford}
    \country{UK}
}
\email{simon.birnbach@cs.ox.ac.uk}

\author{Martin Strohmeier}
\affiliation{%
    \institution{armasuisse Science + Technology}
    \city{Thun}
    \country{Switzerland}
}
\email{martin.strohmeier@ar.admin.ch}

\author{Ivan Martinovic}
\affiliation{%
    \institution{University of Oxford}
    \city{Oxford}
    \country{UK}
}
\email{ivan.martinovic@cs.ox.ac.uk}

\begin{document}


\begin{abstract}

Due to an increase in the availability of cheap off-the-shelf radio hardware, signal spoofing and replay attacks on satellite ground systems have become more accessible than ever.
This is particularly a problem for legacy systems, many of which do not offer cryptographic security and cannot be patched to support novel security measures.

Therefore, in this paper we explore \textit{radio transmitter fingerprinting} in the context of satellite systems.
We introduce the \sysname{} system, proposing novel techniques for authenticating transmissions using characteristics of the transmitter hardware expressed as impairments on the downlinked radio signal.
We look in particular at high sample rate fingerprinting, making device fingerprints difficult to forge without similarly high sample rate transmitting hardware, thus raising the required budget for spoofing and replay attacks.
We also examine the difficulty of this approach with high levels of atmospheric noise and multipath scattering, and analyze potential solutions to this problem.

We focus on the \textit{Iridium} satellite constellation, for which we collected \num{1705202}~messages at a sample rate of \qty{25}{\mega\sample/\second}.
We use this data to train a fingerprinting model consisting of an autoencoder combined with a Siamese neural network, enabling the model to learn an efficient encoding of the message headers that preserves identifying information.

We demonstrate the fingerprinting system's robustness under attack by replaying messages using a Software-Defined Radio,
achieving an Equal Error Rate of~\num{0.120}, and ROC AUC of~\num{0.946}.
Finally, we analyze its stability over time by introducing a time gap between training and testing data, and its extensibility by introducing new transmitters which have not been seen before.
We conclude that our techniques are useful for building fingerprinting systems that are stable over time, can be used immediately with new transmitters without retraining, and provide robustness against spoofing and replay attacks by raising the required budget for attacks.

\end{abstract}

\maketitle

\section{Motivation}\label{sec:motivation}

Recent years have seen a dramatic rise in the availability of cheap radio hardware, particularly Software-Defined Radios (SDRs).
Not only have these devices become more widely available, but their capabilities have increased, with devices like the \textit{HackRF One} (\qty{340}{\usd}, Adafruit) able to transmit and receive in frequencies ranging from \qty{1}{\mega\hertz} to \qty{6}{\giga\hertz}~\cite{greatHackRF2021}.
As a result, the ability to carry out spoofing attacks, once exclusive to large-budget organizations and nation-state actors, is now within reach for even motivated hobbyists.
This poses a particular threat to satellite systems, many of which were built under the assumption that tampering with signals would be prohibitively expensive for the vast majority of attackers.

Spoofing and replay attacks have been widely explored in wireless systems -- attackers equipped with an SDR can overshadow legitimate communications or spoof messages outside normal communication.
Many widely used systems are vulnerable to these attacks, including the ADS-B avionics protocol~\cite{strohmeierSecurity2015}, the LTE telephony system~\cite{lichtmanLTE2016,leeThis2019}, and satellite systems including GPS~\cite{gasparCapture2020}.
Due to the critical nature of satellite systems, it is vital that operators can prevent or detect spoofing attacks to protect the systems and applications that rely on them.

There are a wide range of techniques for detection and prevention of spoofing attacks, the foremost of which is cryptography -- a properly implemented cryptosystem with associated key management provides robust authentication, making spoofing attacks near-impossible.
However, there are a number of reasons why cryptography may not be desirable (or possible) in the context of satellite systems.
Firstly, there are a huge number of legacy satellites currently in orbit.
Many of these do not implement cryptography, and cannot be retrofitted to do so due to their limited onboard processing power.
However, the data collected by these satellites is immensely useful in both scientific and private use cases -- they are used for monitoring forest fires, land usage, population density, flooding, and more~\cite{nasaFirms, esriMap, metaMap, cloudToStreet}.
These satellites are often bespoke designs which would be prohibitively expensive to replace; it is important to ensure systems like these can be used for their entire projected lifespan (and beyond).

There are also a number of satellite systems which were initially built with cryptography, but which have become insecure post-launch due to leaked keys~\cite{xrit-rx} or outdated cryptosystems~\cite{lrit-key-dec}.
Some of these satellites cannot be patched due to a lack of over-the-air update capabilities, so other methods must be used to authenticate their telemetry data.

Finally, some attacks can be carried out without violating any cryptographic properties of the system.
The authors of~\cite{motallebighomiCryptography2022} show that precisely timed message replays can cause Global Navigation Satellite Systems (GNSS) to misreport the location of the receiver to an attacker-specified location.
Since these attacks are carried out by simply introducing delay to messages rather than altering message contents, conventional cryptography does not protect against them.
We still want to be able to detect and prevent these attacks, so we must turn to non-cryptographic techniques for message authentication. In particular, we investigate radio transmitter fingerprinting, in which radio signal characteristics are used to identify the transmitters.
This is achieved by identifying impairments on the signal which are created by small differences in the radio transmitter hardware.
These impairments are unique to transmitters and consistent over time as we will show for the satellite case.

\subsection*{Contributions}

In this paper we present \sysname{}, a novel approach to fingerprinting satellite signals.
We work with signals at a high sample rate to counteract problems with spoofing at lower sample rates.
This makes the techniques more useful in a security context, requiring attackers to use more expensive radio hardware that works at high sample rates in order to successfully impersonate a device -- this excludes a large number of low-budget adversaries.
In doing this we can provide an additional level of confidence in the authenticity of the origin of satellite signals, particularly in systems where cryptography is either unavailable or ineffective.
We verify that the system can detect replay attacks by replaying captured messages using an SDR.

We also use a \textit{Siamese model} -- unlike conventional classifiers, these compare two signals and produce a distance metric representing the likelihood of two messages having been sent from the same transmitter.
This technique enables \textit{one-shot learning}: new transmitters can be introduced without requiring the system to be retrained, and can be used immediately with only a small number of examples.
This is particularly useful in the context of satellites in Low Earth Orbit (LEO), which must be replaced more frequently.

We captured a large high sample rate dataset of real-world data from the \textit{Iridium} satellite constellation.
We use this data to train and test our system, since decoders are readily available and previous works have demonstrated the constellation to be viable for fingerprinting~\cite{oligeriPASTAI2020}.
Our baseline architecture can be used and enhanced for use in deployed satellite systems.
To faciliate this, our full dataset and code have been made freely available.\footnote{The code is available at \url{https://github.com/ssloxford/SatIQ}, and the dataset and model weights at \url{https://zenodo.org/record/8220494} and \url{https://zenodo.org/record/8298532} respectively.}

\section{Background}\label{sec:background}

In this section we discuss key concepts in radio and digital signal processing.
We also explore existing countermeasures to spoofing attacks, and look at existing research in radio fingerprinting in general.
Finally, we discuss related work that focuses on radio fingerprinting of satellites, as well as related work that considers fingerprinting in a security context.

\subsection{Software-Defined Radios}

Software-Defined Radios (SDRs) allow the use of software for signal processing tasks traditionally done using dedicated hardware.
This is achieved by sampling raw signals into a digital form and sending the samples to a computer, where they can be processed further.
This provides significant versatility to signal processing over traditional radio hardware, at the cost of additional processing power.

A concept at the core of digital signal processing is \textit{IQ Sampling}.
By taking a \textit{carrier signal} at a given frequency and sampling the components of the incoming signal that are in phase ($I$) and out of phase (quadrature, $Q$) with the carrier before sampling, the incoming signal is downsampled, shifting the carrier frequency to \qty{0}{\hertz}~\cite{lichtmanIQ2021}.
This significantly reduces the sample rate required.

One benefit of sampling in this way is that samples can be represented as complex numbers, and plotted on the complex plane.
In this representation, distance from the origin represents amplitude and angle from the horizontal axis represents phase relative to the carrier signal.
This makes it a particularly useful representation of \textit{Phase Shift Keying} (PSK), a form of signal modulation which encodes data in the phase of the signal.
In this case, symbols appear as distinct points on the complex plane, producing a \textit{constellation diagram}.

We only see distinct points on the constellation diagram if we sample at the exact symbol rate of the modulation scheme.
If we instead \textit{oversample} the signal by using a significantly higher sample rate, we start to see the points between the symbols, as the transmitter hardware modulates between them.
This can be seen later on in Figure~\ref{fig:iridium-sample} (Section~\ref{sec:data-collection}).
The appearance of this interpolation between points is affected by a number of factors including atmospheric noise and multipath distortion, but also by small variations in the transmitter hardware -- these can be used to fingerprint the transmitter.

\subsection{Spoofing Countermeasures}\label{sec:countermeasures}

In Section~\ref{sec:threat-model} we explore the threat of an attacker equipped with a software-defined radio, looking in particular at spoofing attacks.
When implemented properly, modern cryptographic authentication can solve this problem, but it is also possible to make these systems more secure without the use of cryptography.
As discussed in Section~\ref{sec:motivation}, this is desirable in legacy systems without cryptography, or systems where cryptographic authentication is undesirable or has been compromised.
Additionally, some systems provide open data by design, leaving out cryptography on downlinked communications by choice.

There are a diverse range of approaches for the authentication of downlinked transmissions without the use of cryptography.
These can be partitioned into \textit{data inspection}, \textit{timing analysis}, and \textit{waveform analysis}.

\paragraph{Data Inspection}
The integrity of data can be verified by receiving the same transmission at multiple locations, and comparing the data or its hash between receivers.
This requires an attacker to be physically present at each location in order to successfully carry out spoofing attacks, where they would have previously only needed to be at a single location.
This is particularly useful in cases where there are already large numbers of community-operated ground stations, such as with NASA's Direct Readout Laboratory (\num{168} operated worldwide at the time of writing)~\cite{nasaDirect}.
However, it is likely to be infeasible for smaller organizations to set up multiple ground stations.

\paragraph{Timing Analysis}
These techniques involve looking at the timing of signals in order to verify their legitimacy.
At the simplest level, this could be ensuring the signal is received at a time the satellite was known to be transmitting -- this does not provide much real security.
More advanced techniques in this area include Time Difference of Arrival (TDOA) analysis, looking at the time difference between multiple receivers to verify the transmitter's location against where the satellite is known to be~\cite{jedermann2021orbit}.
This is an effective technique, forcing any potential attackers to be physically present at every ground station, but is once again only feasible for larger organizations capable of operating many ground stations.

\paragraph{Waveform Analysis}
Alongside data and timing analysis, the waveform itself can be inspected to detect attacks.
A spoofed signal is likely to have different properties from the legitimate signal, particularly when spoofing requires overshadowing an ongoing transmission.
These properties include amplitude, SNR, doppler shift, and signal distortion~\cite{maneshDetection2019,mirallesAssessment2020}.
With appropriate radio hardware it is possible to verify these parameters, making attacks more difficult to execute -- the adversary must replicate the measured properties in order to successfully spoof messages.

\textit{Fingerprinting} also falls into this category -- by looking at unique impairments on the raw waveform we can identify the transmitter.
When spoofing or replaying messages the attacker's radio will impart a different fingerprint on the signal, allowing a fingerprinting system to detect when this has occurred.
In order for the adversary to circumvent this system they will need to replicate the fingerprint of the legitimate transmitter.
Depending on how the fingerprinter has been designed, this may raise the required budget to carry out attacks.

\subsection{Fingerprinting}\label{sec:fingerprinting}

Radio fingerprinting is a mature field, with a large base of research looking at a wide range of techniques on many different systems -- \cite{soltaniehReview2020} provides a good overview of existing research.
Fingerprinting techniques can be partitioned into two key areas: \textit{transient fingerprinting} and \textit{steady-state fingerprinting}.

\subsubsection{Transient Fingerprinting}
The \textit{transient} of a radio signal occurs when the transmitter first powers on, or changes power levels following a signal lock.
Various properties of the transient, such as its duration or the number of peaks which occur in the carrier signal, are characteristic to the transmitter and can be used to identify it, if properly extracted.
The majority of historical fingerprinting research makes use of transient analysis, since almost all radio transmission involves a transient, and the transient typically exhibits the same characteristics every time the device powers up.
Much of the work in transient fingerprinting revolves around novel techniques for precisely identifying the start and end of the transient~\cite{hallDetection2003,huangDetection2013}, or processing the transient to extract useful identifying features~\cite{ellisCharacteristics2001}.

Transient fingerprinting has seen some use in security contexts -- in~\cite{rasmussenImplications2007}, transient fingerprinting is used to identify devices even when all other identifying information has been removed, and to detect wormhole and device cloning attacks (these attack types are explored further in Section~\ref{sec:threat-model}).

\subsubsection{Steady-state Fingerprinting}
In contrast to transient fingerprinting looking at a very brief portion of the signal, steady-state fingerprinting instead looks at the modulated portion of the signal.
The features in the steady-state portion of the signal are different from the transient, often looking at how the IQ constellation is affected by hardware impairments.
These impairments include quadrature errors, self-interference, amplitude clipping, and frequency offsets in the various stages of signal modulation (DAC, mixer, filter, upconverter, amplifier)~\cite{foruhandehSpotr2020}.
Additionally, the signal is affected by properties of the wireless channel, including background noise, free space path loss, and multipath distortion, which are likely to change over time as the environment changes.
These wireless channel properties are significantly more prominent in satellite communication, due to the long-distance radio links involved.

There is some variety in the techniques used, including observing features of the constellation at a low sample rate~\cite{oligeriPASTAI2020}, analyzing features in the frequency domain~\cite{kennedyRadio2008}, and looking at high sample rate signals to observe high frequency impairments~\cite{basseyIntrusion2019}.
Machine learning techniques are more commonly used to aid steady-state fingerprinting, particularly in approaches working at a high sample rate, in order to pull out features which may not be immediately obvious.
Manual feature engineering is also used, but is less common than in transient analysis.

The majority of steady-state fingerprinting looks at a single protocol or class of devices at a time, but with the recent rapid increase in machine learning capabilities this is no longer a requirement, and there has been some work into generalizable fingerprinting techniques which do not require retraining to apply to new contexts~\cite{restucciaDeepRadioID2019}.

There are also some interesting techniques extending the concept of fingerprinting -- the authors of~\cite{sankheNo2020} train a convolutional neural network to identify SDRs at \qty{5}{\mega\sample/\second}, then intentionally introduce signal impairments at the transmitter in order to further increase classification accuracy.
This achieves incredible accuracy (greater than \num{0.995}), but fingerprint forgery is not considered in this context.

\subsection{Related Work}\label{sec:related-work}

We have explored the existing body of radio fingerprinting research in the previous section.
However, there is also some research that focuses specifically on satellite fingerprinting and fingerprinting in a security context.

\subsubsection{Satellite Fingerprinting}
Unlike signals from terrestrial devices, satellite signals have to travel hundreds of kilometers through the atmosphere, causing significant signal attenuation and channel noise.
This adds additional challenge to fingerprinting in this context, particularly since many techniques rely on minimal presence of background noise.
There are some works looking at fingerprinting in the presence of noise, either by adding noise to clean signals during model training (effectively training models to remove/ignore the noise)~\cite{tekbasImprovement2004}, or by smoothing out long signals at low sample rates to obtain average symbol positions~\cite{wangRadio2022}.
In the context of satellites it is difficult to obtain signals without noise, and smoothing does not work with high sample rate signals (since important detail is lost) -- we must find other methods of reducing or ignoring noise.
We discuss this further in Section~\ref{sec:system-overview}.

The authors of~\cite{oligeriPASTAI2020} design ``PAST-AI'', analyzing heatmaps of low sample rate transmissions from the \textit{Iridium} constellation to classify satellites.
This technique achieves an accuracy of approximately \num{0.85}, increasing to \num{1.00} for small subsets of the constellation.
Although this technique achieves high accuracy, it is not as useful from a security context -- the classifier works by processing large batches of consecutive messages, making it more difficult to detect individual message spoofing.
Furthermore, fingerprinting at a very low sample rate (\num{1}~sample per symbol) makes fingerprint forgery significantly easier, since the attacker does not need to replicate as many features of the waveform.
The work claims to be able to discriminate SDR-equipped attackers from legitimate satellites by solving the harder problem of discriminating between satellites with the same hardware.
This is true, however, the assumption only holds at the sample rate used by the fingerprinting system.
As PAST-AI operates at \num{1}~sample per symbol, it cannot protect against SDR-based attacks at higher sample rates.

There has also been some fingerprinting work looking at other satellite systems -- the authors of~\cite{foruhandehSpotr2020} make use of manual feature extraction to identify spoofing of GPS localization satellites.
This technique is effective, but manual feature extraction is less likely to transfer easily to other satellites or constellations.

\subsubsection{Security}
Most fingerprinting is done for the purpose of security, preventing spoofing and replay attacks, but some techniques have been specifically proposed to provide better security properties.
For instance, the authors of~\cite{basseyIntrusion2019} address the problem of identifying devices which have not been seen in the training dataset by separating the feature extraction and classification components of the model, using clustering techniques on the extracted features to identify transmitters.
This allows new transmitters to be introduced without retraining the feature extraction component.
We take a similar approach in our work, using an autoencoder to produce the fingerprints and a Siamese model in the place of clustering.
The \sysname{} system is fully described in Section~\ref{sec:system-overview}.

There has also been work assessing attacks on fingerprinting systems -- it has been shown that an arbitrary waveform generator with a sufficiently high sample rate can be used to impersonate devices, fooling fingerprinting systems~\cite{danevAttacks2010}.
Hardware that can achieve these sample rates is prohibitively expensive for the vast majority of adversaries (we discuss budget further in Section~\ref{sec:threat-model}), so we do not consider this to be a major issue -- although our techniques are likely vulnerable to impersonation at a very high sample rate, preventing spoofing below a certain transmitter sample rate is sufficient to exclude the vast majority of lower-budget attackers.
\section{Threat Model}\label{sec:threat-model}

\paragraph{Goal}\label{sec:goal}
In this paper we concentrate on attacks involving spoofing and message replay.
In the case of spoofing, the adversary's goal is to broadcast messages appearing to come from a satellite such that the ground system processes them alongside legitimate messages.
Alternatively, the attacker may delay or advance messages (jamming the original and replaying a recording) to affect timing-based systems such as GPS~\cite{motallebighomiCryptography2022}.
Similarly, they could carry out ``wormhole attacks'', in which messages are captured at one location and tunneled to another location, from where they are broadcast -- this is also effective against GPS and other localization systems.
Unlike spoofing, these attacks can be performed even on signed or encrypted messages, since they do not affect message contents.

\paragraph{Capabilities}\label{sec:capabilities}

The adversary's capabilities and budget will vary depending on the type of adversary.
We can reasonably expect any attacker to have access to an off-the-shelf SDR (as well as the appropriate amplifiers and antennas) enabling them to transmit messages within the vicinity of a single ground station.
An equipment setup similar to the one given in Section~\ref{sec:data-collection} could be used to carry out attacks, costing approximately \qty{6600}{\usd}.
By using a cheaper SDR with a lower sample rate and a different antenna (with a suitable amplifier), the budget can be significantly lowered to approximately \qty{500}{\usd}.\footnote{For example, a HackRF One can be used as the radio (\qty{340}{\usd}, Adafruit), with a suitable power amplifier and passive antenna.}
Alternatively, with a greater budget the attacker could afford a more powerful amplifier enabling messages to be broadcast from a greater distance, or multiple copies of the hardware to attack multiple ground stations simultaneously.

It has been demonstrated in~\cite{danevAttacks2010} that device fingerprinting is vulnerable to signal replay attacks, provided the attacker has access to a high-end arbitrary waveform generator capable of transmitting signals with a sufficiently high sample rate.\footnote{The authors approached a reputable manufacturer, and received a quote for approximately \qty{125000}{\usd} at academic institution rates. We therefore conservatively assume that in the best case this hardware would cost no less than \qty{60000}{\usd}.}
For this reason we do not consider attackers with nation-state level capabilities, since they are always capable of purchasing hardware that can fool a fingerprinting system.

In this work we are particularly interested in lower-budget attacks; a simple spoofing or replay attack with a cheap SDR can have a potentially devastating effect on improperly secured satellite systems.
Through robust high-sample-rate device fingerprinting, we aim to defeat these attacks by making it impossible to forge the fingerprint on spoofed signals using only cheap COTS radio hardware.
In doing so, we increase the budget of attacks such that they can no longer be carried out by lower-budget attackers.

Of course, we do not expect fingerprinting techniques to be able to prevent all attacks; such a goal is unrealistic for this scope, requiring robust cryptography.
As discussed in Section~\ref{sec:motivation}, there are a large number of satellites with outdated or no cryptography and no capability for in-flight patching, and we argue that it is still crucial to secure these satellites as well as possible despite this, so they can still be used to serve customers and for the advancement of science.
We discuss in more detail in Section~\ref{sec:evaluation} how the threat model changes with the introduction of robust signal fingerprinting, addressing remaining concerns and potential avenues for future research.

We also do not consider attacks from within a compromised system, in which compromised satellites are communicating with other devices in the network.
This further means that spoofing attacks launched from satellites within the constellation that our system protects are out of scope for this work.
Existing systems such as PAST-AI~\cite{oligeriPASTAI2020} are already well-suited to tackle this type of attacker.
However, as these existing systems are based on low-sample rate data, they cannot protect against SDR-based attacks -- the focus of this work.

\begin{figure}
    \centering
    \includegraphics[width=\linewidth]{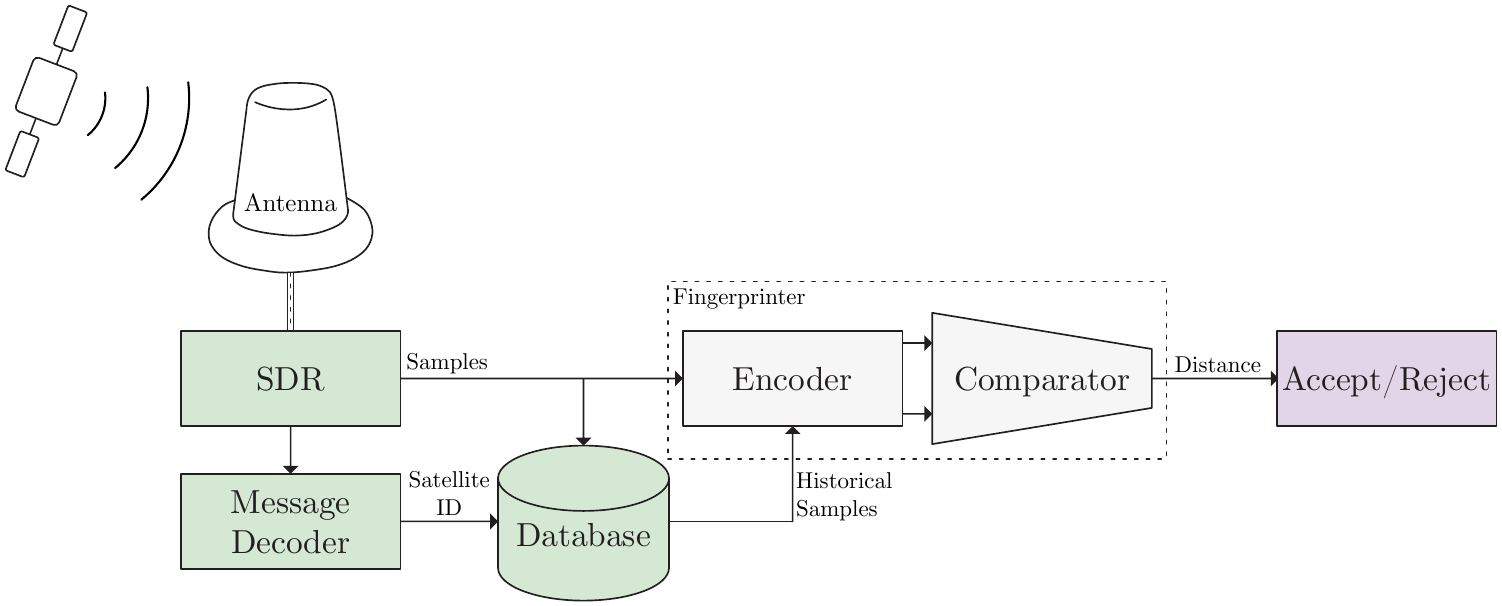}
    \caption{An overview of the end-to-end fingerprinting process used by \sysname{}. Satellite signals are received, decoded, processed into fingerprints, and compared to historical fingerprints to determine a distance metric, which is used to accept or reject the message.}
    \label{fig:fingerprinting-overview}
    \Description[A diagram showing the hardware and software steps used by SatIQ.]{A diagram showing the hardware and software steps used by SatIQ. A satellite transmits signals to an antenna connected to an SDR. The message is decoded and the satellite ID alongside the raw samples are stored in a database. The samples and historical samples from the database are passed to the fingerprinter (an encoder and comparator) which produces a distance score, which is used to accept or reject the sample.}
\end{figure}

\begin{figure}
    \centering
    \includegraphics[width=.7\linewidth]{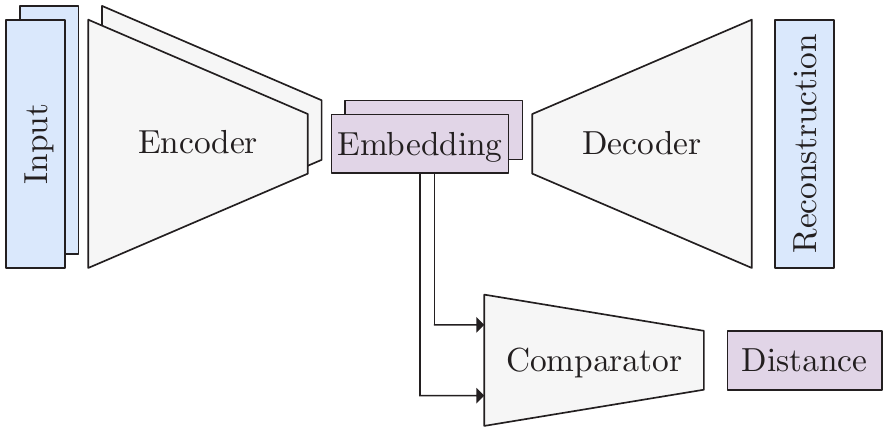}
    \caption{An overview of the Siamese autoencoder architecture used in \sysname{}'s fingerprinting model. Two inputs are passed into the encoder with identical weights, and the encodings are compared using the comparator (angular distance) to generate a distance metric.}
    \label{fig:ae-siamese-overview}
    \Description[A diagram of a Siamese autoencoder.]{A diagram of a Siamese autoencoder, in which two inputs are passed through the same encoder to produce two embeddings. These are passed through a decoder to produce a reconstruction, as well as a comparator to produce a distance.}
\end{figure}

We are primarily concerned with attackers overshadowing messages on the downlink -- whilst attacks on the uplink are possible, they have not been extensively explored due to the greater hardware cost of a suitable amplifier and directional dish.
Furthermore, device fingerprinting on the space segment is currently infeasible, requiring large amounts of computational power, and cannot be carried out aboard the legacy satellites with which we are primarily concerned.
Further work may consider fingerprinting the uplink of legacy satellite systems by capturing signals in-transit, but this is a very limited use case and is out of scope for this research.

\section{System Design}\label{sec:system-overview}

Our system comprises two primary components: data collection and the \sysname{} machine learning fingerprinting system.
A representation of the end-to-end system can be seen in Figure~\ref{fig:fingerprinting-overview} -- messages are collected by an SDR and decoded into message contents, and the raw samples are ran through an encoder network (explained below), and compared against known example messages from the same transmitter.
This produces a distance metric -- low distances indicate the message is likely to be legitimate.
This is used to accept or reject the message.

All code and datasets have been made openly available.
The source code can be found at \url{https://github.com/ssloxford/SatIQ}, and the dataset and model weights at \url{https://zenodo.org/record/8220494} and \url{https://zenodo.org/record/8298532} respectively.

\begin{figure}
    \centering
    \includegraphics[width=.8\linewidth]{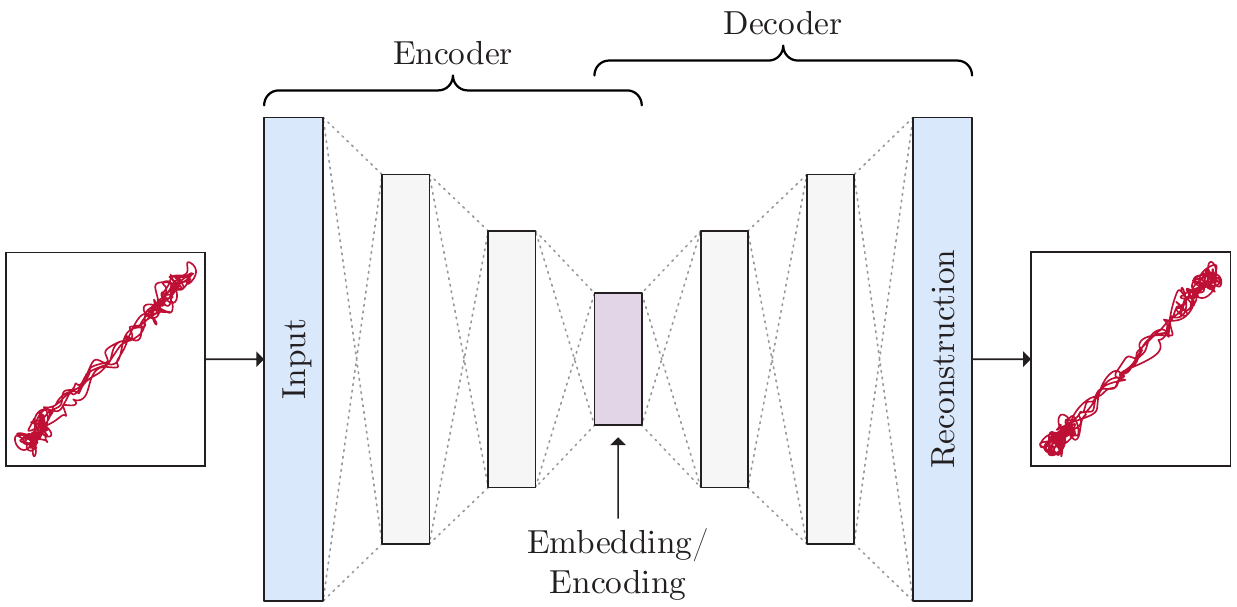}
    \caption{The layout of an autoencoder. Input is passed through an encoder to produce a low-dimensional encoding, and then through a decoder to produce output of the same dimension as the input. The model is trained to reconstruct the input as best as possible.}
    \label{fig:autoencoder}
    \Description[A diagram showing an autoencoder's layout.]{A diagram showing an autoencoder's layout, with an input waveform passed through an encoder with successively smaller layers to produce a small encoding. This is passed through a decoder with successively bigger layers to produce a reconstruction of the input.}
\end{figure}

\subsection{Design Decisions}\label{sec:design-decisions}

Our fingerprinting model, illustrated in Figure~\ref{fig:ae-siamese-overview}, uses an autoencoder combined with a Siamese model to compare two input waveforms and produce a distance metric between the two inputs.
In the following, we describe autoencoders and Siamese networks in general, and we explain why we chose them as the basis for our fingerprinting system.

\paragraph{Autoencoders}\label{sec:autoencoders}

\begin{figure*}
    \centering
    \includegraphics[width=\linewidth]{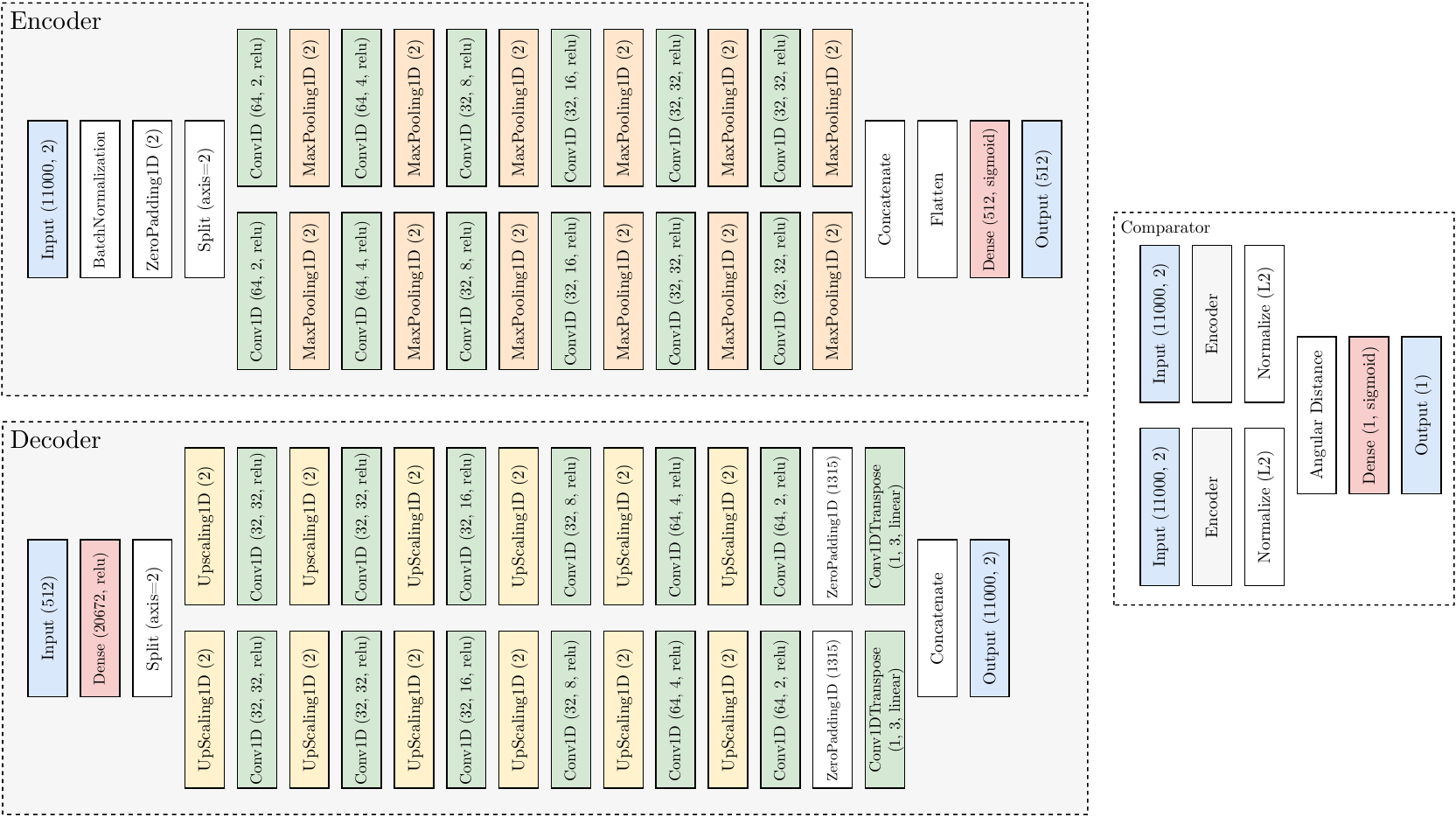}
    \caption{The layers of the Siamese neural network used in \sysname{}. The encoder uses separate convolutional and max-pooling layers for the $I$ and $Q$ portions of the signal, before producing a final embedding using a dense layer. Similarly, the decoder uses separate convolutional and upsampling layers. The comparator uses two copies of the encoder with identical weights, computing a difference score between the outputs.}
    \label{fig:ae-siamese}
    \Description[A diagram showing the layers in the SatIQ neural network.]{A diagram showing the layers in the SatIQ neural network. The encoder is composed of convolutional and max pooling layers. The decoder is composed of convolutional and upscaling layers. The comparator passes two inputs through the encoder, takes the angular distance between the two results, and passes the result through a dense layer to produce its output.}
\end{figure*}

An autoencoder is a type of neural network which is used to learn an efficient encoding of data.
This is achieved by simultaneously training an \textit{encoder} and \textit{decoder}, validating the accuracy of the encoding by comparing the output of the decoder to the input (reconstruction accuracy).
The output of the encoder is restricted in size, thus forcing data to pass through a \textit{bottleneck} at this portion of the network.
This forces the model to discard less important information, producing an efficient encoding.
This technique is particularly useful for dimensionality reduction, since the encoder produces an embedding of the input in a significantly lower-dimensional space.
The layout of an autoencoder can be seen in Figure~\ref{fig:autoencoder}.

\paragraph{Siamese Neural Networks}\label{sec:siamese-neural-networks}

Siamese neural networks are designed to be effective for one-shot classification. To this end, they generate a similarity score between two inputs~\cite{chiccoSiamese2021}.
This is achieved by passing each input through the same ``encoder'' network to generate an embedding of the inputs in some feature space, followed by a comparison function to generate a distance metric in the feature space -- the lower the distance, the more similar the samples are.
The weights of the encoder network are shared between the two inputs.

We chose this approach due to some of its advantages over a simple classifier, particularly in the context of fingerprinting:
\begin{itemize}
    \item The number of classes is not fixed -- new classes can be introduced after training by comparing to examples from the new class.
    \item The one-shot (or in some cases few-shot) nature of the model means a new class can be identified using only a very small number of examples.
    \item The distance threshold can be raised to increase the acceptance rate of legitimate messages at the cost of increased false positives (or vice versa), granting fine-grained control of the level of protection granted by fingerprinting.
\end{itemize}

Past work has shown these models to be effective in a wide range of use cases, particularly in classification problems with huge numbers of classes such as malware detection or gait recognition~\cite{zhuMultiLoss2020,zhangSiamese2016}.
The architecture has also been demonstrated to be successful at detecting spoofing and replay attacks on other systems, such as face recognition and voice biometrics~\cite{peiPersonSpecific2023,sriskandarajaDeep2018}.
Siamese networks have also seen use in radio systems, shown to be effective in areas such as automatic modulation classification~\cite{maoAttentive2021}.
Finally, some research has shown promise in using Siamese networks for fingerprinting radio transmitters~\cite{morge-rolletSiamese2020,langfordRobust2019}.

Our work builds upon these in a number of key aspects.
Firstly, we deal with a more difficult scenario than the majority of terrestrial fingerprinting cases -- the great distance of satellite transmitters introduces large amounts of atmospheric noise and multipath distortion, which dwarf the hardware impairments on the signal.
Secondly, we consider in particular the security implications of radio fingerprinting, assessing the level of protection against spoofing and replay attacks granted by our approach, and the expected budget required to circumvent these techniques.
Finally, we use an autoencoder alongside the Siamese network to provide encodings that better capture meaningful features in the input -- prior work has shown this to be effective in areas such as signature verification, but to our knowledge we are the first to apply this architecture to radio fingerprinting~\cite{ahrabianUsage2019}.

In a non-adversarial classification setting, simple classifiers typically exhibit higher performance than Siamese networks, since they partition the entire input space into fixed categories.
However, this is less effective in adversarial settings, since malicious transmitters will still be given a legitimate label.
Our results in Section~\ref{sec:evaluation} show that Siamese architectures are particularly effective at detecting replay attacks, even with the levels of noise observed in satellite signals.

Siamese models can also require more data during training, since they must learn a distance metric that works for all transmitters.
Furthermore, independently of the model architecture, greater levels of noise mean more training data is required.
Our results show that our dataset was sufficient for this work, but more data may yield further performance improvements.

\subsection{Fingerprinting Model}\label{sec:fingerprinting-model}
\begin{figure}
    \centering
    \includegraphics[width=\linewidth]{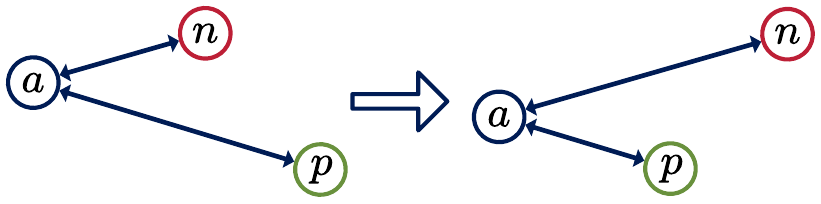}
    \caption{An illustration of the triplet loss function. This function takes an anchor $a$, a positive sample $p$ of the same class as the anchor, and a negative sample $n$ of a different class. Optimizing this loss function minimizes the distance between the anchor and positive samples, and maximizes the distance between the anchor and negative samples.}
    \label{fig:triplet-loss}
    \Description[An illustration of triplet loss.]{An illustration of triplet loss, with circles representing the anchor, positive, and negative samples. After training, the positive sample is closer to the anchor and the negative sample is further away.}
    \vspace{-0.2em}
\end{figure}

The Siamese network of \sysname{} uses the encoder portion of the autoencoder to produce embeddings of two inputs, which are then compared to one another using an angular distance function.
A triplet loss term encourages the model to produce embeddings that are close to one another for messages from the same transmitter, and different for messages from different transmitters:
\begin{align*}
    d(u,v) &= 1 - \frac{u \cdot v}{\lVert u \rVert \lVert v \rVert}\\
    L(a,p,n) &= \max\left(d(a,p) - d(a,n) + \alpha, 0\right)
\end{align*}
This takes an anchor $a$, a positive example $p$ belonging to the same class as the anchor, and a negative example $n$ of a different class, and encourages the distance from $a$ to $p$ to be less than $a$ to $n$.
A margin $\alpha$ ensures that effort is not wasted on optimizing triplets for which this is already the case.
This is illustrated in Figure~\ref{fig:triplet-loss}.

Convolutional layers are used in the encoder to enable identification of position-invariant features and reduce the overall size of the model.
For the comparator, we compute the angular distance between the embeddings -- this tends to work better than L2 distance for high-dimensional data.
A diagram of the fingerprinting model architecture used in this paper can be seen in Figure~\ref{fig:ae-siamese}.

The inclusion of convolutional layers reduces the model's size, and allows it to extract position-invariant features from the waveform.
We also use separate layers for the in-phase ($I$) and quadrature ($Q$) portions of the signal -- although the components are tightly coupled to one another, we find that they express different features and the model is able to perform better when the two are separated.
Following the convolutional layers, we concatenate the outputs and flatten, before using a fully connected layer to reduce the output to the correct size.
The decoder uses a very similar architecture to the encoder, composed of alternating upsampling and convolutional layers.

\subsection{Data Collection}\label{sec:data-collection}

For a fingerprinting model to be effective, a good dataset is essential.
Community projects such as \textit{SatNOGS} (an open-source network of ground stations~\cite{satnogs}) provide data from a wide range of satellites.
However, in this paper we collect our own data for a couple of important reasons: firstly, collecting our own data enables us to capture at a significantly higher sample rate than existing datasets, providing a good foundation for a fingerprinting system.
We also capture signals from a specific constellation rather than a collection of individual satellites; by doing this, we ensure that the signal modulation and protocol does not vary between messages.
We can therefore guarantee that the message header is always the same between messages, so differences in the captured waveform will be caused only by hardware differences and channel noise -- the contents are always identical at the bit level.
This gives us a consistent baseline upon which a fingerprinting model can be built.

We focus on the \textit{Iridium} constellation, used in telecommunications.
This constellation has a number of useful properties:
\begin{itemize}
    \item The constellation contains a large number of satellites (\num{66}, each with \num{48} transmitters~\cite{veenemanIridium2021}), providing sufficient variety within the dataset;
    \item The transmitter hardware on each satellite is identical, so a fingerprinter will need to distinguish satellites only through differences introduced at time of manufacture, rather than distinguishing between entirely different components;
    \item The communication protocols are known and well documented~\cite{veenemanIridium2021}, so no reverse engineering is required;
    \item Downlinked transmissions can be received using cheap and widely available COTS hardware.
\end{itemize}

Other constellations are available which satisfy some of these properties, most notably \textit{Starlink} and Planet Labs' \textit{Dove} constellation; transmissions from these are more difficult to capture, but could be useful for future evaluations of the versatility of our system.
Furthermore, existing research in fingerprinting also uses Iridium satellites~\cite{oligeriPASTAI2020}, providing evidence that some fingerprinting is plausible in this context and providing a baseline to which fingerprinting systems can be compared.
We argue that testing on Iridium is sufficient to demonstrate and validate the effectiveness of \sysname{}, as the properties of the signal used by \sysname{} are not constellation-specific.

\begin{figure}
    \centering
    \includegraphics[width=.8\linewidth]{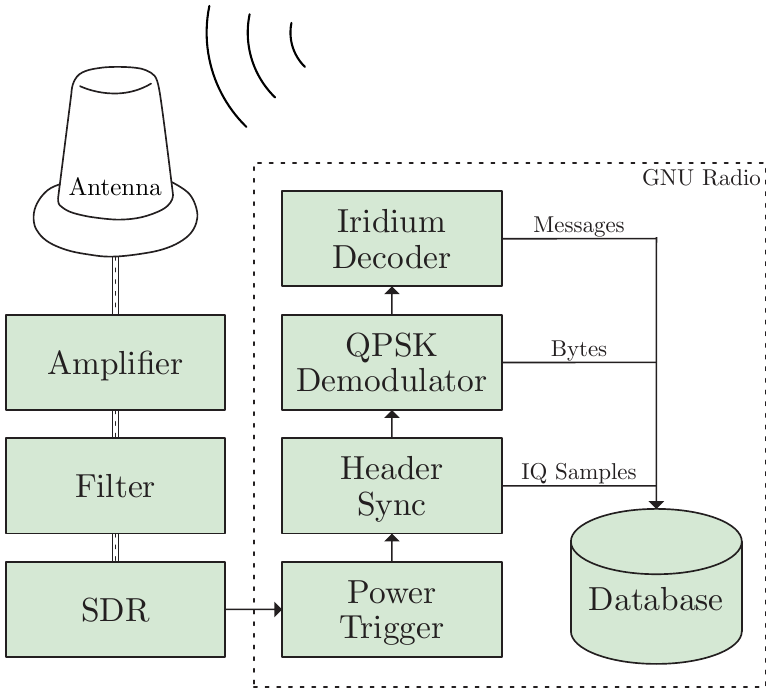}
    \caption{An overview of the data collection system. Signals are captured by the SDR and sent to GNURadio for processing. Raw IQ samples, demodulated bytes, and fully decoded message data are all sent to a database.}
    \label{fig:data-collection}
    \Description[A diagram showing how the hardware and software components are connected in SatIQ.]{A diagram showing how the hardware and software components are connected in SatIQ. The following components are connected in sequence: antenna, amplifier, filter, SDR, power trigger, header sync, QPSK demodulator, Iridium decoder, database. The header sync component sends IQ samples to the database, the QPSK demodulator sends bytes, and the Iridium decoder sends messages. Everything after (but not including) the SDR is in software, using GNU radio.}
\end{figure}

For the data collection itself, we use the following hardware:\footnote{Prices are as recorded on 2023-04-21, and may not reflect current prices.}
\begin{itemize}
    \item Iridium Beam active antenna (\qty{1245}{\usd}, Beam Communications)
    \item Mini-Circuits ZKL-33ULN-S+ low-noise amplifier (\qty{209}{\usd}, Mini-Circuits)
    \item NooElec DC block (\qty{20}{\usd}, Amazon)
    \item Mini-Circuits VBF-1560+, \qtyrange[range-phrase=--,range-units=single]{1500}{1620}{\mega\hertz} band pass filter (\qty{44}{\usd}, Mini-Circuits)
    \item USRP N210 SDR (\qty{3354}{\usd}, Ettus Research)
    \item UBX 40 USRP daughterboard (\qty{1732}{\usd}, Ettus Research)
\end{itemize}
The SDR is connected to a computer running \textit{GNU Radio}, a software library to aid digital signal processing.
We use components from the \textit{gr-iridium} library, created and maintained by the Chaos Computer Club M{\"u}nchen e.V., to demodulate and decode messages~\cite{schneiderGNU2022}.
Figure~\ref{fig:data-collection} illustrates our full data collection and processing pipeline.

Iridium downlink messages have a number of different message types, one of which is the \textit{Iridium Ring Alert} (IRA) message -- this is the only type of message we collect.
These messages contain diagnostic information about the satellite, including the satellite ID, beam ID (identifier of the current transmitter), position, and altitude.
They also contain the anonymous identifiers of subscribers currently receiving incoming paging calls.
The messages are openly broadcast and do not contain any personally identifiable information, so we can decode and collect them without raising any ethical concerns.

\begin{figure}
    \centering
    \includegraphics[width=.8\linewidth]{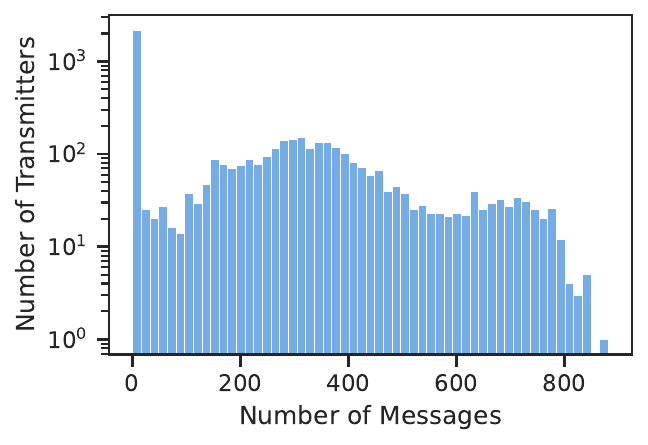}
    \caption{The distribution of the number of Iridium messages received per transmitter.}
    \label{fig:iridium-histogram}
    \Description[A histogram showing the number of transmitters in the y axis, and number of messages in the x axis.]{A histogram showing the number of transmitters in the y axis, and number of messages in the x axis. There is a spike at 0, and the rest of the distribution is flatter, tailing off at around 800.}
\end{figure}

We save the raw IQ samples from the message headers in a database, alongside demodulated bytes and decoded message contents.
This gives us a consistent dataset of raw message headers which can be used for fingerprinting, using the decoded messages to label the data.

Our antenna was located on the roof of a building in Oxford, UK, giving it a good view of the sky in all directions.
We collected messages at \qty{25}{\mega\sample/\second}, giving us an oversampling rate of \num{1000} (Iridium messages have a symbol rate of \num{25000} symbols per second).
This allows us to capture the high-frequency features characteristic to a transmitter for effective fingerprinting.
In total we collected \num{1010464}~messages over \num{23}~days.
For each transmitter we collected up to \num{872} messages, with a mean of \num{203}~messages per transmitter -- the distribution of message count per transmitter can be seen in Figure~\ref{fig:iridium-histogram}.
We also collected a further \num{694738}~messages over \num{17}~days immediately following the training data collection -- this data will be used for evaluation in Section~\ref{sec:evaluation}.

\begin{figure}
    \centering
    \includegraphics[width=\linewidth]{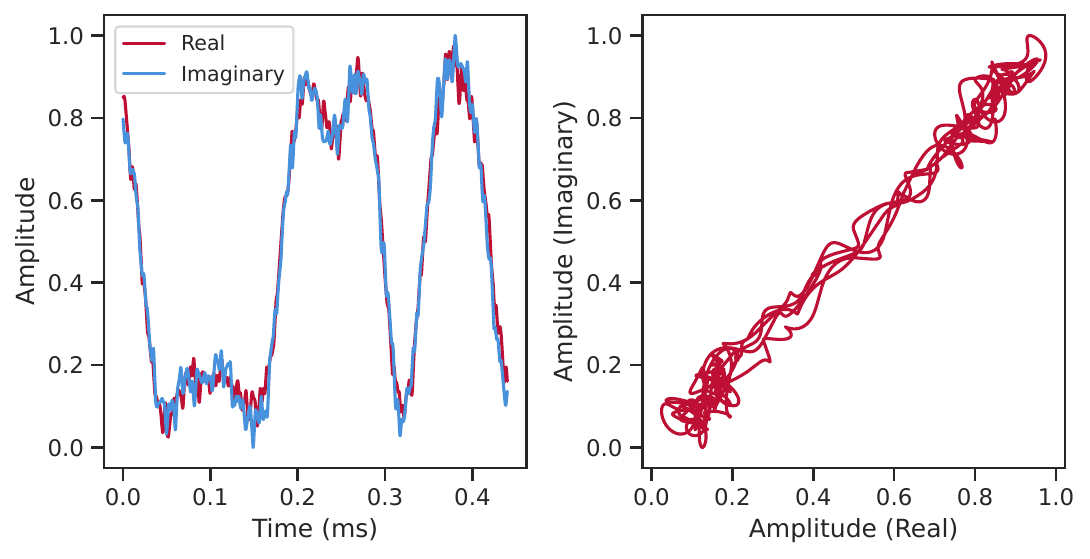}
    \caption{A message header received from an Iridium satellite, shown in the time domain (left) and as a constellation plot (right).}
    \label{fig:iridium-sample}
    \Description[Two plots of an Iridium message header.]{Two plots of an Iridium message header. The left-hand plot is in the time domain (amplitude against time), and the right-hand plot is a constellation plot (in-phase component against quadrature component). Both plots show the signal interpolating between two symbols in opposite corners.}
\end{figure}

An example of the collected data can be seen in Figure~\ref{fig:iridium-sample}.
We can see the signal encodes \num{8}~QPSK symbols, corresponding to the bit sequence for the Iridium message header: \texttt{11 00 00 11 11 00 11 00}.
However, unlike a constellation plot at \num{1}~sample per symbol, we can see the movement between the two symbols and the impairment on the signal.
It is clear that there is significant impairment; this is likely caused by a combination of channel noise, multipath distortion, and hardware characteristics of the transmitter.
Our goal is to isolate the last of these, ignoring the channel noise and other characteristics -- these will be the same between transmitters, and not useful in this context.

\subsubsection{Data Preprocessing}\label{sec:data-preprocessing}

We opt to use a minimal amount of preprocessing to avoid destroying data which might be useful for fingerprinting.
On top of the band-pass filtering and phase synchronization performed by \textit{gr-iridium}, we scale each waveform in the dataset so values range between \complexnum{-1-j} and \complexnum{1+j}.
This makes visualization easier, and removes magnitude as an additional factor the model needs to learn.
We also remove all messages which do not decode as valid IRA messages -- although this removes a large number of messages, it ensures that all data is labeled and eliminates the noisiest messages which do not properly decode.
This leaves us with messages that are the most likely to contain meaningful identifying factors.

Finally, we shuffle the data and process it into ``TFRecord'' files -- this format is optimized for use in TensorFlow, storing data as raw protocol buffers.
This ensures we can load the data efficiently and minimize RAM usage by reading mostly from disk.

\subsubsection{Dataset Construction}\label{sec:dataset-construction}

After data preprocessing, we split the data into training, validation, and testing sets, and use a generator to produce batches of examples.
We set aside \num{50000}~messages for each of the validation and testing datasets, resulting in a training:validation:testing split of $90{:}5{:}5$.
This ensures we have sufficient data to evaluate the model without significantly reducing the size of the training dataset.

Since we are using a triplet loss function, we produce batches of inputs with \num{32}~messages (\num{4}~from each of \num{8}~different transmitters) -- this ensures the loss function can select a large number of triplets in each batch.

\subsection{Model Training}\label{sec:model-training}

Model training was performed on a machine with the following hardware:
\begin{itemize}
    \item 2$\times$ Intel Xeon E5-2660 v3 CPU (20 threads, \qty{3.30}{\giga\hertz})
    \item Nvidia TITAN RTX GPU (\qty{1770}{\mega\hertz}, \qty{24}{\giga\byte} VRAM)
    \item \qty{128}{\giga\byte} DDR4 RAM (\qty{2133}{\mega\hertz})
\end{itemize}

On this hardware, up to 3 model configurations can be trained simultaneously.
All models are trained for \num{100} epochs, taking \numrange[range-phrase=--,range-units=single]{12}{24}~hours.

Running a trained model takes significantly fewer resources, taking a fraction of a second to verify messages -- this makes it possible to run \sysname{} in real-time, verifying messages as they arrive.
Furthermore, much less RAM is required, as the training dataset does not need to be loaded.
Hardware requirements are discussed further in Section~\ref{sec:implementation}.

\subsection{Model Optimization}\label{sec:results}

\begin{figure}
    \centering
    \begin{subfigure}{0.49\linewidth}
        \includegraphics[width=\textwidth]{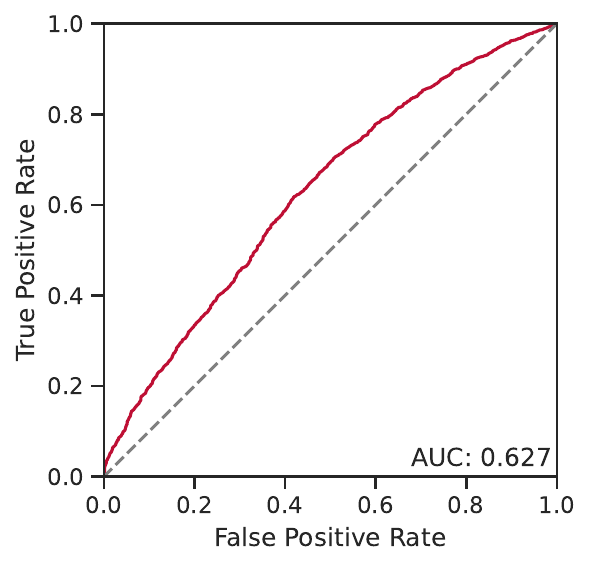}
        \caption{ROC curve}
        \label{fig:results-base-roc}
        \Description[ROC curve of SatIQ.]{ROC curve of SatIQ, plotting true positive rate against false positive rate. The curve is above the diagonal, and has an area under curve (AUC) of 0.627.}
    \end{subfigure}
    \begin{subfigure}{0.475\linewidth}
        \includegraphics[width=\textwidth]{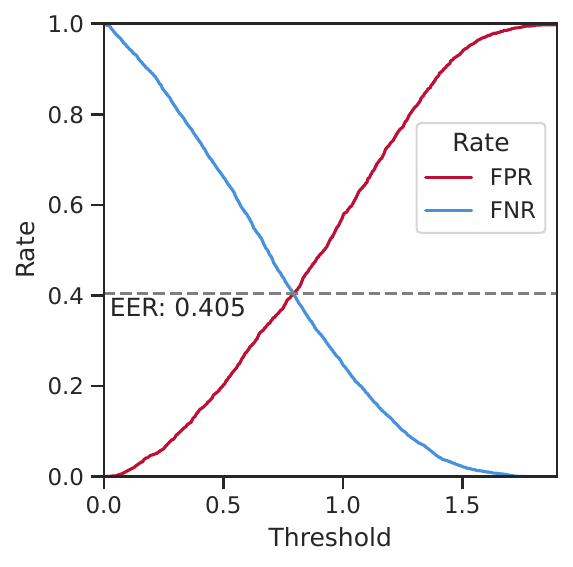}
        \caption{FPR against FNR}
        \label{fig:results-base-fpr-fnr}
        \Description[Plotting false positive rate against false negative rate.]{Plotting false positive rate against false negative rate, for the same model as the other subfigure. The lines cross at an error rate of 0.405.}
    \end{subfigure}
    \caption{The performance of the base \sysname{} model, evaluated by comparing messages from different pairs of satellites without analyzing the attack case.}
    \label{fig:results-base}
\end{figure}

Figure~\ref{fig:results-base} shows the base performance of \sysname{}.
In this section, models are trained and tested on the problem of differentiating satellites in the dataset, in order to achieve better performance later on when identifying adversaries in an attack scenario.
Results in this section are primarily to assess relative performance and fine-tune the system.
When assessing performance, we focus on two key metrics:
\begin{itemize}
    \item \textit{Equal Error Rate (EER):} the error rate when the false positive rate (FPR) and false negative rate (FNR) are equal.
    \item \textit{Area Under Curve (AUC):} the area under the Receiver Operating Characteristic (ROC) curve, obtained by plotting the true positive rate (TPR) against the FPR. This can be intuitively thought of as the probability that the system can distinguish between two inputs of different classes~\cite{narkhedeUnderstanding2018}.
\end{itemize}
In computing each of these, we vary the distance threshold below which we accept two messages as being from the same transmitter.
By raising this threshold, we accept a greater number of legitimate messages, but open the system up to easier spoofing attacks.
Conversely, by lowering the threshold the system is made more secure, but legitimate messages are less likely to be accepted as such.

\begin{figure}
    \centering
    \includegraphics[width=.7\linewidth]{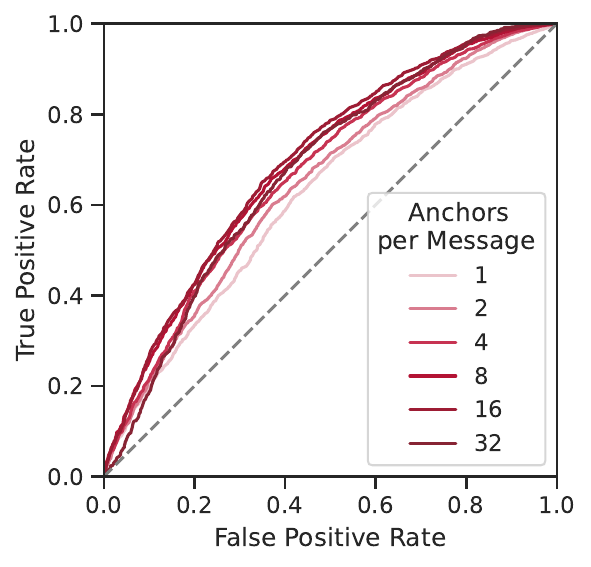}
    \caption{ROC curves as we compare each incoming message to a larger number of anchors, taking the mean distance.}
    \label{fig:roc-mean}
    \Description[Multiple ROC curves stacked on top of each other.]{Multiple ROC curves stacked on top of each other, showing a small increase in performance as the number of anchors per message increases from 1 to 32.}
\end{figure}

\begin{figure}
    \centering
    \includegraphics[width=.7\linewidth]{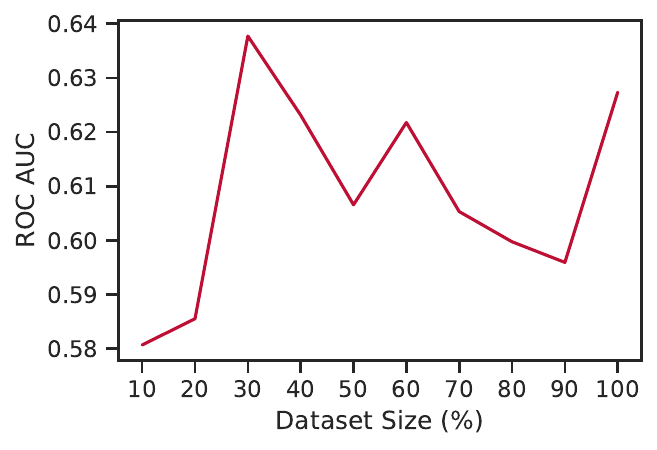}
    \caption{The AUC performance of the fingerprinting model as the size of the training dataset is changed. In each instance, the messages with the least noise are kept.}
    \label{fig:reduced-dataset}
    \Description[A line graph.]{A line graph plotting model AUC against the size of the dataset used to train the model. Performance is low for the smallest datasets, and there is a peak at 30 percent.}
\end{figure}

Our base model has an AUC of \num{0.627} and EER of \num{0.405}.
This is sufficient performance to demonstrate the feasibility of our techniques, particularly in the more difficult case of distinguishing satellite transmitters with identical hardware.
In Section~\ref{sec:evaluation} we go on to demonstrate that performance is significantly better in a replay attack scenario, confirming its usefulness in a security context.
For the rest of this section we continue to assess performance on the original (non-attacked) dataset.

\subsubsection{Multiple anchors}\label{sec:multiple-anchors}

To improve performance over the base system, we can compare each incoming message to a larger number of ``anchors'' (known messages from that transmitter), taking the mean distance between the message and each anchor in the embedding space.
The results of this are shown in Figure~\ref{fig:roc-mean}.
By taking \num{16}~anchors for each incoming message, we can achieve an EER of~\num{0.350} and AUC of~\num{0.698} -- a significant improvement!
In practice, this can be implemented by saving a larger set of messages from each known transmitter, or by comparing multiple consecutive messages to the same set of anchors.
Both of these techniques are practical -- our observations suggest that during an Iridium phone call or web connection approximately \num{11}~packets are exchanged per second, so an attacker will need to spoof many packets to have a meaningful impact on the victim.
Such an attack would certainly be picked up by \sysname{}, even if multiple consecutive messages are compared.
We explore attack scenarios further in Section~\ref{sec:evaluation}.

\subsubsection{Reduced dataset}

As satellites pass overhead, their downlinked signals are subject to different amounts of attenuation due to changing distances and different thicknesses of the atmosphere along the direct path between the satellite and the observer.
As a result, our dataset contains messages of differing amplitudes and levels of noise.
Although all messages are normalized before use, this is likely to have an effect on performance; it will be more difficult to extract identifying information from signals with high levels of background noise.

We can try to mitigate this by filtering the dataset, removing the messages with the greatest levels of noise.
For this experiment we produced 9 datasets with different percentages of messages from the original dataset removed.
Figure~\ref{fig:reduced-dataset} shows the AUC of models trained on these smaller datasets.
We can see that performance is greatest with the dataset composed of the least noisy~\num{30}\%, with an AUC of~\num{0.638} and EER of~\num{0.407} -- below this, the negative effect of the dataset's small size exceeds the benefit provided by using cleaner inputs.

\begin{figure}
    \centering
    \includegraphics[width=\linewidth]{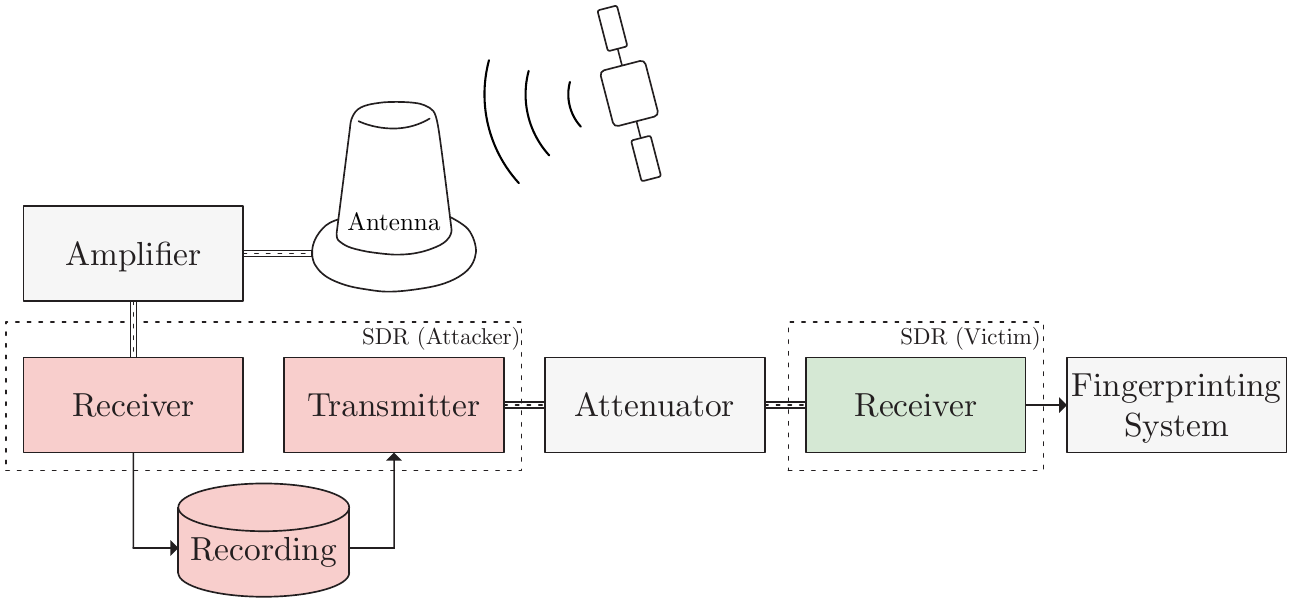}
    \caption{Hardware setup for the replay attacks. Raw samples are captured using an SDR, then replayed directly into the fingerprinting system's SDR over a cable.}
    \label{fig:replay-attack}
    \Description[A flowchart showing the hardware setup for replay attacks.]{A flowchart showing the hardware setup for replay attacks. The following components are connected in order: antenna, amplifier, receiver, recording, transmitter, attenuator, receiver, fingerprinting system. The first receiver, recording, and transmitter represent the attacker's hardware, and the second receiver represents the victim's hardware.}
\end{figure}

Care must be taken when using this approach in practice, as operators will need to decide whether messages above this noise threshold are accepted or rejected by default, or wait for a suitably clean message before the transmitter can be authenticated.
If multiple satellites are visible at once, the ground system can connect to the satellite with the best SNR -- this is how real-world satellite systems work in practice, connecting to satellites that are more directly overhead.
For the remainder of this paper, we use the full dataset.

\section{Evaluation}\label{sec:evaluation}

In the previous section we designed and trained the \sysname{} system to distinguish between different legitimate satellite transmitters, without considering a real-world attack scenario.
In this section we evaluate \sysname{} for its robustness against replay attacks, extensibility to new satellite transmitters, and stability over time.
We also discuss further techniques for assessing these types of model, including transferrability to new satellite constellations.
Finally, we consider the practical questions surrounding a real-world deployment of \sysname{}.

\begin{table}
    \centering
    \sisetup{table-number-alignment=center}
    \begin{tabular}{S[table-format=0.3]S[table-format=0.3]S[table-format=0.3]}
        \toprule
        {TPR} & {FPR} & {Threshold} \\
        \midrule
        0.999 & 0.978 & 1.343 \\
        0.990 & 0.818 & 1.198 \\
        0.950 & 0.307 & 1.055 \\
        0.900 & 0.150 & 0.977 \\
        0.861 & 0.100 & 0.932 \\
        0.805 & 0.050 & 0.885 \\
        0.672 & 0.010 & 0.795 \\
        0.424 & 0.000 & 0.665 \\
        \bottomrule
    \end{tabular}
    \caption{True positive (true accept) rates and false positive (false accept) rates for key threshold values, tested on replayed messages. Messages are tested against 32 anchors, and the mean distance is taken.}
    \label{tab:acceptance-thresholds}
\end{table}

\subsection{Security}

We first evaluate the security properties of \sysname{}, assessing its performance under an attack scenario.
The simplest such scenario involves merely swapping the identifiers of transmitters in our existing dataset.
This models an attack from within a compromised system, in which an attacker has gained control of a satellite currently in use -- this matches our training scenario and results.
As discussed in Section~\ref{sec:threat-model}, this scenario is somewhat unrealistic, and not our primary concern.

The more interesting case is to evaluate \sysname{}'s robustness under real-world replay attacks.
We can realistically evaluate this by performing replay attacks over-the-wire.
These do not perfectly model real-world spoofing and replay attacks, but any concessions made are in the attacker's favor -- for instance, they do not have to account for path loss or background noise as they would in a radio setting.
A similar result could be achieved through the use of an RF-shielded box, but this has the potential to introduce further impairments through reflections and other effects.
Our experimental setup is as shown in Figure~\ref{fig:replay-attack}.
We first capture Iridium messages at \qty{25}{\mega\sample/\second}, saving the raw IQ samples to a file -- this provides us with a dataset of samples identical to what the SDR would normally receive.
We then replay these samples over a wire connected to the ``victim'' SDR, feeding the captured messages into the fingerprinting system.

\begin{figure}
    \centering
    \includegraphics[width=.7\linewidth]{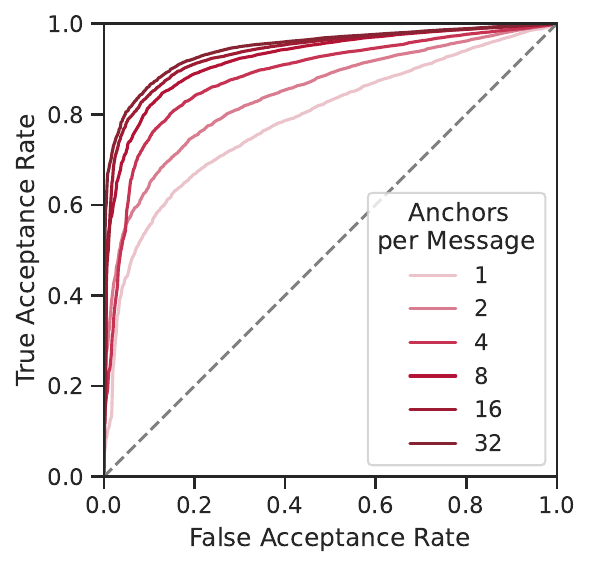}
    \caption{ROC curves showing the system's performance when detecting replayed messages. Performance is significantly better than our training results, achieving a maximum AUC of 0.946.}
    \label{fig:roc-replay-mean}
    \Description[Multiple ROC curves stacked on top of each other.]{Multiple ROC curves stacked on top of each other, showing a significant increase in performance as the number of anchors per message increases from 1 to 32.}
\end{figure}

For each replayed message, we take a number of ``known good'' messages from the same transmitter from our testing dataset.
We randomly select a number of these messages to be our anchors using a ``shuffle split'' strategy.
We compare the anchors to the replayed messages to obtain the false positive rate, and to the other known good messages to get the true positive rate.
The results of this experiment are shown in Figure~\ref{fig:roc-replay-mean}.
We can see that \sysname{} performs significantly better in this scenario, with a base AUC of~\num{0.788}.
When we compare each message to \num{32} anchors this performance increases even further, with an AUC of~\num{0.946} and an EER of~\num{0.120}!
This indicates that the attacker's SDR has introduced its own fingerprint, distorting the message and altering its features.

Furthermore, this performance is good enough to deploy in a real-world system -- by adjusting the acceptance threshold we can achieve a high true acceptance rate while minimizing the number of spoofed messages that are accepted.
These results are summarized in Table~\ref{tab:acceptance-thresholds}.
By setting the threshold such that \num{90}\% of legitimate messages are accepted, we accept only \num{15}\% of the attacker's messages.
This performance is good enough to use in a real-world setting, particularly if we continuously fingerprint messages over the course of a communication session, taking the average acceptance rate over time as an indicator of attack -- in order to have a meaningful impact the attacker will need to spoof multiple messages, which significantly raises the likelihood of detection.

This attack scenario assumes a well-equipped adversary with access to a high-end SDR, and eliminates all the difficulties of over-the-air replay attacks.
Despite all these concessions, we are still able to detect the attack in the majority of cases.
With an even higher budget (to transmit at an even higher sample rate) and careful effort to eliminate noise introduced by the radio, it will certainly be possible to circumvent this system~\cite{danevAttacks2010}, but our results show that it will take a concerted effort to do so -- simple message replay is not enough.
We can therefore exclude a large proportion of attackers with all but the highest budgets, granting a real-world security benefit to ground systems.

\subsection{Extensibility}

\begin{figure}
    \centering
    \includegraphics[width=.7\linewidth]{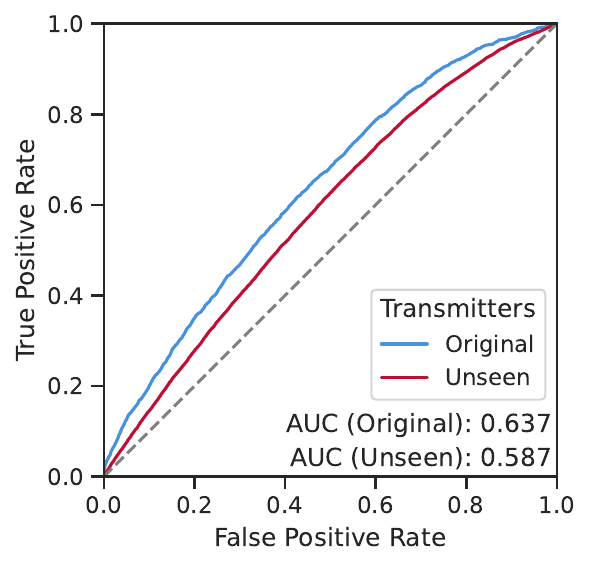}
    \caption{ROC curve showing the system's performance on transmitters it has never seen before.}
    \label{fig:roc-anchor-removed-transmitters}
    \Description[Two ROC curves stacked on top of each other.]{Two ROC curves stacked on top of each other, showing the original model's performance, and its performance on transmitters it has not seen before. There is a slight decrease in AUC for unseen transmitters, going from 0.637 to 0.587.}
\end{figure}

Next we look at how well \sysname{} performs on transmitters it has never seen before.
This is of particular importance in satellite constellations, where satellites may need to be replaced at any time, and we want to minimize time and effort spent retraining the model.
To test \sysname{}'s extensibility, we trained the fingerprinting model on a dataset with \num{50} transmitters removed.
The results of this analysis are shown in Figure~\ref{fig:roc-anchor-removed-transmitters}.
As expected, the base performance roughly matches our original model's testing performance -- an AUC of~\num{0.637} when testing against \num{16}~anchors.
When tested on the \num{50} transmitters removed from its training data, the performance drops slightly to an AUC of~\num{0.587}.
With this level of performance we can continue to use the system with slightly reduced accuracy -- in the context of satellites, this is better than existing classifier-based systems, which require full retraining each time a satellite is launched or replaced.
With further training and tuning, it may be possible to completely eliminate even this small performance drop, making a system that can be used indefinitely as transmitters are added and replaced.

It may also be possible to transfer a trained \sysname{} model to a completely new satellite constellation, since many of the signal impairments will be common between hardware configurations.
With a small amount of retraining, similar performance might be achieved across a wide range of satellite systems.
Such analysis is beyond the scope of this paper, as it will require a new dataset for the second constellation.

\subsection{Time Stability}\label{sec:time-stability}

Finally, we evaluate \sysname{}'s stability over time.
Existing works in radio fingerprinting observe a decrease in performance when there is a time gap between the training and testing data~\cite{hamdaouiDeepLearningBased2022,al-shawabkaExposing2020}.
This is thought to be caused by changes in the conditions of the wireless channel over time.
To assess the effect of this on \sysname{}, we test our model on the additional dataset collected after the training data, comprising \num{694738}~messages over \num{17}~days.

\begin{figure}
    \centering
    \includegraphics[width=.8\linewidth]{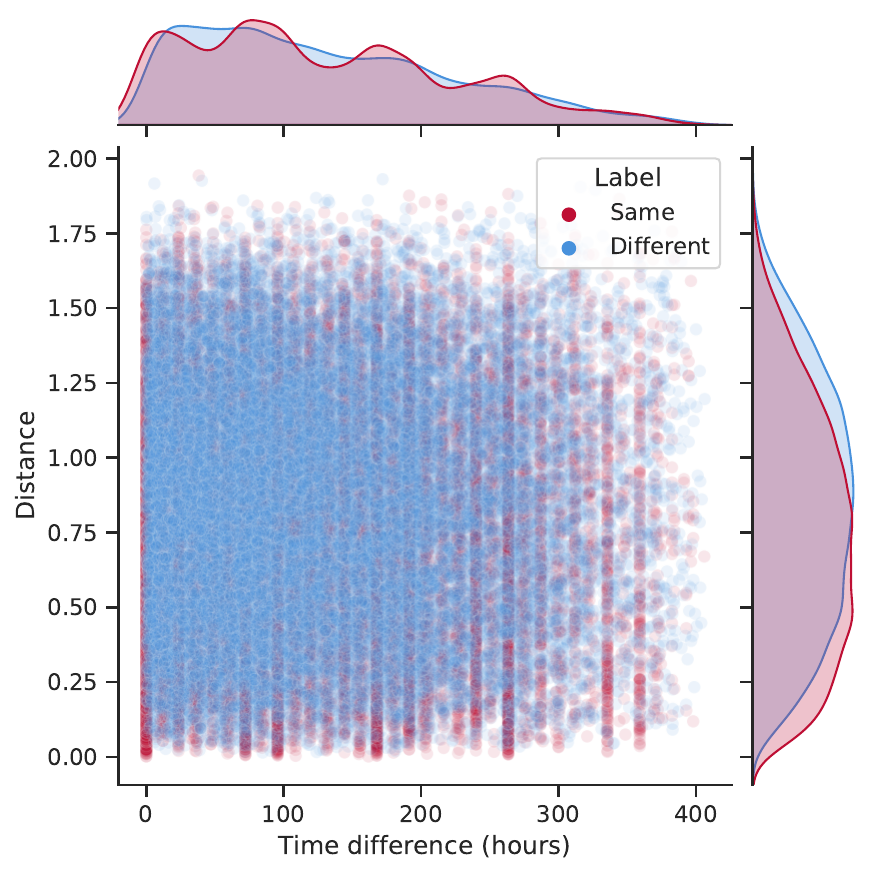}
    \caption{The relationship between the time difference between messages and their distance in the embedding space, separated by messages from the same transmitter and different transmitters.}
    \label{fig:timediffs}
    \Description[A joint plot containing a scatter plot and kernel density estimates.]{A joint plot containing a scatter plot and kernel density estimates, plotting distance in the embedding space against time difference between pairs of samples. The distribution is noisy, and it is hard to discern a correlation.}
\end{figure}

Figure~\ref{fig:timediffs} shows how the distance between two messages in embedding space is affected by the time difference between the two messages.
There is not a strong correlation between these two factors -- further analysis is required to determine the exact relationship between time difference and performance.

\begin{figure}
    \centering
    \includegraphics[width=.7\linewidth]{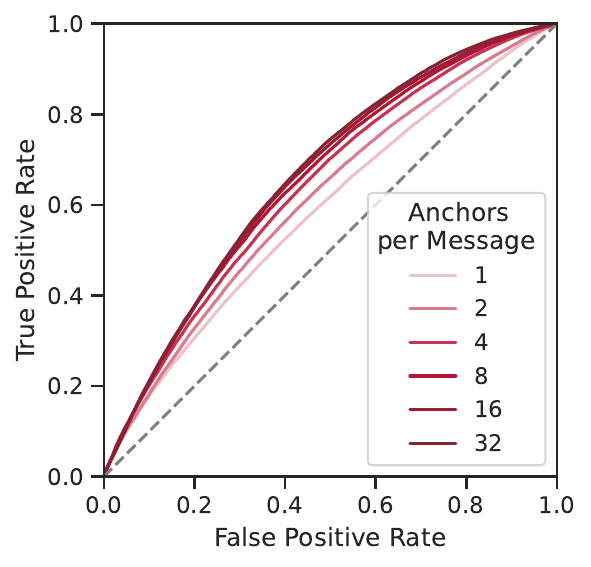}
    \caption{ROC curves depicting performance on new data collected after the training data, comparing each incoming message against multiple anchors. At 32 anchors per message, an AUC of 0.659 is achieved.}
    \label{fig:roc-anchor-timediff-fresh-multiple}
    \Description[Multiple ROC curves stacked on top of each other.]{Multiple ROC curves stacked on top of each other, showing similar performance on new data as in the original plot. As before, performance increases as the number of anchors per message increases from 1 to 32.}
\end{figure}

We tested our trained model on the final \num{24}~hours of the dataset, comprising \num{43601} messages with a \num{16}~day time gap between this data and our training dataset.
As expected, we observe a moderate decrease in performance when comparing messages to anchors from our original test dataset, with an AUC of \num{0.580}.
However, if we use fresh anchors from the new dataset the overall AUC improves to \num{0.615} -- this is only a small decrease from the training results, which compare older anchors to old messages.
This performance can be improved further by testing against multiple anchors as in Section~\ref{sec:multiple-anchors} -- Figure~\ref{fig:roc-anchor-timediff-fresh-multiple} illustrates this, showing that a testing AUC of \num{0.659} can be achieved with an EER of~\num{0.385} (close to our original AUC and EER of \num{0.698} and \num{0.350}).

We therefore propose that a deployed system based on \sysname{} replaces its anchor messages (used for testing incoming messages) from time to time, ensuring freshness and maximizing performance.
However, caution must be exercised to ensure the anchors are not inadvertently replaced with spoofed messages, as this would entirely circumvent the security of the fingerprinting system.
Secure refreshing of the anchors could be achieved by periodically verifying messages through other means -- for instance, by precisely measuring angle of arrival.

\subsection{Deployment Considerations}\label{sec:implementation}

In this section we discuss what it would look like to deploy \sysname{} in a real-world system.
As we mention in Section~\ref{sec:model-training}, a trained \sysname{} model can be run in real-time, verifying messages as they arrive.
We validate this by running the model on lightweight hardware.
Our data collection observed roughly \num{0.5} ring alert messages per second, or \numrange{6}{7} messages per second across all message types.
This gives us approximately \qty{150}{\milli\second} to validate each incoming message.
On the Intel Xeon CPU used for training (disabling the GPU), the model takes \qty{4.85}{\second} to validate a batch of \num{543} messages (\qty{8.9}{\milli\second} per message), with a maximum RAM usage of \qty{1.7}{\giga\byte} -- this is more than fast enough to validate all incoming messages.

We also ran \sysname{} on a Raspberry Pi 4B single-board computer, with \qty{4}{\giga\byte} of RAM and a \qty{1.8}{\giga\hertz} quad-core processor.
On this hardware, it took \qty{93.25}{\second} to validate the same \num{543} messages (\qty{171}{\milli\second} per message), using \qty{2.0}{\giga\byte} of RAM.
This demonstrates that \sysname{} can run on lightweight hardware -- although slightly too slow to validate all messages in real time, this can certainly validate incoming ring alert messages, and could run even faster with the addition of a dedicated AI accelerator.

We must also consider how deployed devices can capture high sample rate message fingerprints.
Our current setup requires a moderately powerful computer to perform the necessary high sample rate message synchronization and decoding, but this can be reduced significantly with the use of a dedicated FPGA demodulator.
Combined with the good performance on low-power CPUs shown above, it will be possible to integrate \sysname{} into even lightweight devices -- however, some instances may require the addition of hardware.

Additionally, it is important to decide what action is taken when a (potentially legitimate) message is rejected by \sysname{}.
The action taken depends on the level of security required.
For some systems it may be sufficient to simply notify the user, but still accept the message, providing increased awareness of potential attacks without risking any impact to service through rejected legitimate messages.
Alternatively, higher security applications may choose to reject all messages flagged by \sysname{}, accepting the increased packet loss in exchange for greater security against spoofing and replay attacks.
It may be possible for some systems to request message retransmission in this case, to ensure messages are still delivered -- this may occur at the transport or application layer, or systems may be modified to allow retransmission requests (if this has not already been implemented).

Finally, we must consider the logistics of such a deployment.
As discussed in Section~\ref{sec:time-stability}, we propose that \sysname{}-based systems periodically refresh their anchor messages to maximize performance.
An additional benefit of this approach is that no centralized database of fingerprints is required, significantly reducing the complexity of deploying \sysname{} -- there is no need to gather or maintain a database of fingerprints, and devices do not need to connect to a server to update their databases or verify incoming messages.
\sysname{}-based systems can therefore also be developed by either the satellite operator or the receiver manufacturer, or by a third party building devices to integrate with existing ground systems.

\section{Future Work}\label{sec:future-work}

Our research has revealed several promising avenues of future research.
Firstly, it is likely that better performance can be achieved through fine-tuning and a larger dataset, using the same base model architecture as \sysname{}.
In particular, a dataset collected using multiple receiver configurations would likely produce a system that transfers more readily to different ground systems.

Alongside collecting additional data, it will also be useful to implement some of the other evaluation techniques, particularly in assessing the extent to which a trained system can be transferred to another constellation, and what degree of retraining is required.
Furthermore, it would be particularly beneficial to standardize the other analyses such that they could be applied to other fingerprinting techniques.
We currently have no method of empirically comparing fingerprinting techniques to each other in terms of their security properties -- a standard suite of tests would remedy this.
We consider a comparison between fingerprinting systems to be out of scope for this paper, as each fingerprinting technique relies on different kinds of data.
For example, PAST-AI uses full messages rather than just message headers~\cite{oligeriPASTAI2020}, and transient fingerprinting requires precise transient synchronization.

Another promising area of research would be to assess the effectiveness of fingerprinting in conjunction with other methods of spoofing detection, such as assessing SNR or distortion.
Multiple fingerprinting methods could also be used in concert, providing even greater effectiveness than any model alone.
However, some methods are likely to learn the same characteristics as each other (providing no mutual benefit), so a full analysis is needed in order to understand which methods are effective together.

Finally, it would be useful to assess the effectiveness of fingerprinting in systems which already have some amount of authentication.
For instance, such an analysis could evaluate fingerprinting as a preventative measure against GNSS message delay/advancement attacks~\cite{motallebighomiCryptography2022}.
This would demonstrate that fingerprinting is not just useful in legacy systems, but has concrete usefulness even in new satellite systems.

\section{Conclusion}\label{sec:conclusion}

In this paper we have contributed new methods towards radio signal fingerprinting in the context of satellite transmitters, providing novel techniques which can be used to build high sample rate fingerprinting systems.
We have succeeded in demonstrating that satellite signals at high sample rates contain sufficient identifying information, and confirmed that our techniques combining autoencoders and Siamese models are feasible for fingerprinting.

We have also provided a large dataset of captured message headers from Iridium satellites, which can be used for further research and testing in satellite transmitter fingerprinting.
This lays good groundwork for future research in this area.

We have laid out a clear path for future experimental work, both in increasing the accuracy of existing models, and in evaluating the performance of fingerprinting systems from a security context.
Finally, we have discussed the potential for the extensibility of trained fingerprinting models, working across multiple constellations with little or no retraining.

Our work shows that high sample rate fingerprinting is possible and can improve the security against spoofing and replay attacks for the vast majority of low-budget attackers significantly.
This will enable us to continue to use legacy satellite systems and their data with an increased degree of confidence in their integrity.
We verified this by demonstrating that our system can successfully detect replay attacks even in the scenario of a powerful attacker sending messages via a wired channel, achieving an Equal Error Rate of~\num{0.120} and ROC AUC of~\num{0.946}.

\section*{Acknowledgments}
We would like to thank armasuisse Science + Technology for their support during this work.
Joshua and Sebastian were supported by the Engineering and Physical Sciences Research Council (EPSRC).

\balance
\printbibliography

\end{document}